\documentclass[trackchanges]{aastex701}
\usepackage{multirow}
\usepackage{amsmath}
\usepackage{booktabs}

\newcommand{\gmrtext}{$\mathrm{GMRT_{ext}}$}

\begin{document}

\title{Low Frequency Radio Observation of the Dark PeVatron LHAASO J2108+5157 using uGMRT - II}

\author[orcid=0000-0003-4544-2585,sname='Mahanta']{Gunindra Krishna Mahanta}
\affiliation{Astrophysical Sciences Division, Bhabha Atomic Research Centre, Mumbai, 400085, Maharashtra, India}
\affiliation{Homi Bhabha National Institute, Anushaktinagar, Mumbai, 400094, Maharashtra, India}
\email[show]{guninmohantaba@gmail.com, gmahanta@barc.gov.in}  

\author[sname='Roy']{Subhashis Roy}
\affiliation{National Centre for Radio Astrophysics, Pune University Campus, Post Bag 3, Ganeshkhind, Pune, 411007, Maharashtra, India}
\email{}  

\author[sname='Godambe']{Sagar Godambe}
\affiliation{Astrophysical Sciences Division, Bhabha Atomic Research Centre, Mumbai, 400085, Maharashtra, India}
\email{}  

\author[sname='Ghosal']{Bitan Ghosal}
\affiliation{Astrophysical Sciences Division, Bhabha Atomic Research Centre, Mumbai, 400085, Maharashtra, India}
\email{}  

\author[sname='Bhatt']{Nilay Bhatt}
\affiliation{Astrophysical Sciences Division, Bhabha Atomic Research Centre, Mumbai, 400085, Maharashtra, India}
\email{}

\author[sname='Bhattacharyya']{Subir Bhattacharyya}
\affiliation{Astrophysical Sciences Division, Bhabha Atomic Research Centre, Mumbai, 400085, Maharashtra, India}
\affiliation{Homi Bhabha National Institute, Anushaktinagar, Mumbai, 400094, Maharashtra, India}
\email[show]{subirb@barc.gov.in}


\begin{abstract}
We present deep radio observations of the PeVatron candidate LHAASO J2108+5157 using the upgraded Giant Metrewave Radio Telescope (uGMRT). Our observations reveal a persistent double-lobed radio source (\gmrtext) within the vicinity of the $\geq$100 TeV $\gamma$-ray emission region, with consistent morphology across 400 MHz, 650 MHz, and 1.2 GHz. No alternative particle accelerators, such as supernova remnants or pulsar wind nebulae, are identified in the region, making \gmrtext\ the most plausible counterpart candidate. We measure the flux and spatially resolved spectra of the source using multi-band radio data. The morphology remains stable across all bands, while a comparison of non-contemporaneous observations suggests possible flux variability. From the measured spectra of the lobes the inclination angle of the jet is estimated to be $86.57^{\circ} \pm 1.15^{\circ}$. The multi-wavelength spectral energy distribution is modeled reasonably well with a lepto-hadronic scenario.

\end{abstract}


\keywords{\uat{High energy astrophysics}{739} ---\uat{Radio jets}{1347}---\uat{Astronomy image processing}{2306}---\uat{Astronomy data reduction}{1861}}


\section{Introduction}

The origin of cosmic rays (CRs) remains a long-standing problem in high-energy (HE) astrophysics. It is well established that the CR spectrum, with energy ranging from $\sim$10 GeV to $\sim$100 EeV, can be described by a broken power law (PL) with two prominent spectral breaks. The first break, occurring at $\sim$3 PeV, is known as the \textit{knee} of the CR spectrum. It is widely believed that CRs up to the knee are predominantly produced within our Galaxy. Consequently, astrophysical objects capable of accelerating particles up to PeV energies are of particular interest, as they represent the extreme Galactic accelerators. Such sources are commonly referred to as PeVatrons.

Particles accelerated to PeV energies can interact with the surrounding medium and produce $\sim$100 TeV $\gamma$-rays. In the hadronic scenario, high-energy protons undergo $p$–$p$ (or $p$–$\gamma$, depending on the environment) interactions, leading to the production of neutral and charged pions. The decay of neutral pions subsequently gives rise to ultra-high-energy (UHE) $\gamma$-rays. In contrast, in the leptonic scenario, UHE emission is produced via inverse Compton scattering of soft photons by relativistic electrons. Therefore, the detection of $\geq$ 100 TeV $\gamma$-rays is considered a key signature of both hadronic and leptonic PeVatrons.

In recent years, photons with energies $\geq$100 TeV have been detected by several observatories, including HAWC, Tibet AS-$\gamma$, and MAGIC \citep{Amenomori_2019,Abeysekara_2020, Albert_2020,MAGIC_2020}. However, the era of PeVatron physics has been significantly advanced with the detection of dozons of galactic PeVatron by the Large High Altitude Air Shower Observatory (LHAASO) \citep{Cao_2021}. Among the sources reported by LHAASO, several are associated with known astrophysical objects such as supernova remnants (SNRs), pulsar wind nebulae (PWNe), and massive star clusters.

However LHAASO J2108+5157 is one of such sources which lacks any clear association with known astrophysical counterparts. Moreover, no emission from this source has been identified across the electromagnetic spectrum, making it a particularly intriguing object. Recently LHAASO have released its first catalog, in which position of the source LHAASO J2108+5157 is updated and renamed the source as 1LHAASO J2108+5153u. We observed the source region covered by LHAASO field of view (fov) using upgraded Giant Metrewave Radio Telescope (uGMRT) in search of any possible counterpart of the LHAASO source. Observation at 650 MHz with uGMRT revealed a bipolar jet-core like structure of the source which can be a possible counterpart of the LHAASO dark PeVatron. Results were already reported by \cite{Mahanta_2025} (hereafter Paper I). To complete the spectral and morphological study of the source, we further observed the source with uGMRT in 400 MHz and 1.2 GHz frequencies. Here we report the results of our new observations and discuss its consequences.

\section{About LHAASO J2108+5157}

LHAASO J2108+5157 was first reported by \cite{Cao_2021} as part of the discovery of dozons of Galactic PeVatron candidates. The source position was initially determined as RA(J2000) = $317.95^{\circ}$ and Dec(J2000) = $51.95^{\circ}$. The spectrum of the source measured between 20 and 500 TeV, is well described by a single power law,
$
\frac{dN}{dE}= (1.59\pm 0.35_{\rm stat}) \times 10^{-15}\left(\frac{E}{20,\mathrm{TeV}}\right)^{-2.83\pm0.18}\,\mathrm{TeV^{-1}\,cm^{-2}\,s^{-1}}
$.
The source is reported to be a point like object; however, extended morphology couldn't be excluded due to limited statistics and uncertainties in the KM2A point-spread function (PSF). The upper limit on entension is reported to be $\sim$ $0.26^{\circ}$.

Subsequent observations with both LHAASO instruments—KM2A and WCDA—refined the source position to RA(J2000) = $317.10^{\circ}$, Dec(J2000) = $51.90^{\circ}$ (KM2A) and RA(J2000) = $316.83^{\circ}$, Dec(J2000) = $51.95^{\circ}$ (WCDA). In the first LHAASO catalog, the source extension is reported as $r_{39} = 0.14^{\circ} \pm 0.03^{\circ}$ by KM2A and $0.19^{\circ} \pm 0.02^{\circ}$ by WCDA.

Despite extensive follow-up, no low-energy counterpart has been identified. The source is observed by LST1 for nearly 91 hours, yielding only an upper limit \citep{Abe_2023,Abe_2024_pos}. The source region is also observed by HAWC, VERITAS, and XMM-Newton \citep{Adams_2025}. While HAWC reported 7$\sigma$ detection of the source, analysis of 35 hours of good quality VERITAS data yeilds only upper limit. XMM-Newton also could not detect any X-ray counterpart of the source. Presence of a Fermi source named, 4FGL J2108+5155 is also reported in literature \citep{Cao_discovery_2021, Abe_2023, Adams_2025}, but considering the extension of the Fermi object and its spectral behaviour, association of the Fermi source with the PeVatron source couldn't be established. 
CO line study of the source region by CfA 1.2 m telescope indicates the presence of molecular clouds in this region \citep{Dame_2001, Miville_2017}. \cite{Cao_discovery_2021} reported the presence of two molecular cloud viz. [MML2017]2870 and [MML2017]4607. Later CO spectral line observation by Nobayama radio telescope results in the detection of three new molecular cloud [FKT-MC]2022, [FTK-MC]2108, [FTK-MC]HS within the source region \citep{Fuente_2023a, Fuente_2023b}. Presence of molecular cloud in the source region hints towards a hadronic emission scenario.

Motivated by the absence of confirmed low-frequency counterparts, we carried out observations of the source region with uGMRT at 650 MHz. The resulting radio map reveals a double-lobed structure reminiscent of a microquasar or radio galaxy (see Paper I). Result of 650 MHz observation of the source region is summarized in section \ref{sec:band4_obs}.  Although the 650 MHz emission revals a double-lobed structure, the nature of the source could not be identified from that observation. In this work observe the source region in band 3 and band 5 frequency range. Result of the our observation is summarize in section \ref{sec:band3_5_obs}.

\section{Observations and Data Analysis} \label{sec:data_analysis}

The double-lobed source identified in our band 4 observations (hereafter \gmrtext) was followed up with the upgraded Giant Metrewave Radio Telescope (uGMRT) in Bands 3 and 5. band 3 observations were conducted on 2025 June 07 and 24, with a total observation time of 16 hr, while band 5 observations were carried out on 2025 July 21 for 10 hr. The phase center was set at the position of \gmrtext.

The band 3 observations cover the 300–500 MHz frequency range with 2048 spectral channels over a bandwidth of 200 MHz. The band 5 observations span 1060–1460 MHz with 2048 channels over a 400 MHz bandwidth, both in continuum imaging mode.

Flux density and bandpass calibration were performed using 3C286 and 3C48 as primary calibrator, observed at the beginning and end of each observing run. Phase calibration was carried out using the nearby sources 2038+513 (band 3) and 2202+422 (band 5), observed for 5 min after every 30 min scan on the target.

Data reduction and imaging were performed using the Common Astronomy Software Applications (CASA)\footnote{https://casa.nrao.edu/} package following standard procedures. For details of data analysis see Paper I.

\section{Results and Discussion} \label{sec:Results}

\subsection{Band 4 overview}\label{sec:band4_obs}

The morphology and location of \gmrtext\ in 650 MHz emission map is already reported in Paper I (see Fig. 2 of \cite{Mahanta_2025}). The source exhibits a double-lobed structure, suggestive of either a microquasar or a radio galaxy. The double lobed morphology of \gmrtext\ is typically associated with the collimated outflows generally observed in systems such as radio galaxies and microquasars. While an extragalactic radio galaxy origin cannot be excluded, if the GMRT source is physically associated with LHAASO J2108+5157 and is responsible for the observed PeV emission, an extragalactic radio-galaxy interpretation would face significant challenges, as $\gamma$-rays at such energies are expected to suffer substantial $\gamma$–$\gamma$ attenuation over cosmological distances. In contrast, a Galactic microquasar scenario naturally provides a site for sustained particle acceleration to PeV energies, making it a plausible counterpart to the PeVatron candidate. It has been recently argued that the microquasars can be the accelerator of PeV particles \citep{Kaci_2025,Cao_2025,Abaroa_2024,Zhang_2025}. Recent observations showed that few UHE $\gamma$-ray sources observed by HESS and LHAASO without clear counterparts have double-lobed radio structures within the UHE emission region. This emerging pattern is illustrated in various cases, for example

\begin{itemize}
\item \textbf{TeV~J2032+4130:} TeV J2032+4130, is a dark PeVatron candidate, which have no associated counterpart. The source is observed with WSRT telescope at 1400 MHz frequency. The 1400 MHz radio map reveal a dual lob structure in the region of the Dark PeVatron \citep{Butt_2008}. The lack of optical or X-ray counterparts, along with extended radio lobes, suggests that particle acceleration is driven by jets in a faint or hidden system.
\item \textbf{V4641~Sgr:} V4641 Sge is a well established galactic PeVatron. A recent reanalysis of VLA data of the source region reveal a dual lob structure \citep{Marti_2026}. With the detection a double lob source author also associate observed $\sim$ 100 TeV emission with that source.

\item \textbf{1LHAASO~J0343+5254u:} This PeVatron candidate, reported in the first LHAASO catalog, has been observed with LOFAR at 144 MHz. The radio map reveals a double-lobed structure spatially coincident with the high-energy source \citep{Elder_2026}. The detection of such morphology in another LHAASO field further supports the possibility that extended, lobe-like radio emission traces sites of efficient particle acceleration.

\end{itemize}

Taken together, these systems point toward a potential observational connection between double-lobed radio structures and UHE $\gamma$-ray emission. While the current sample size remains limited, the repeated occurrence of such morphologies in PeVatron environments argues against a purely coincidental association.

In this context, the double-lobed structure of \gmrtext\ becomes particularly significant. As one of the few such sources identified within the vicinity of a PeVatron lacking confirmed counterparts, it provides a compelling target for detailed spectral and morphological investigation. Hence, we carried out follow-up observations with uGMRT in bands 3 and 5 to probe its broadband radio properties. The results are discussed below.

\subsection{Results of GMRT band 3 and band 5 Observations}\label{sec:band3_5_obs}

\begin{figure}
    \centering
    \includegraphics[width=\linewidth]{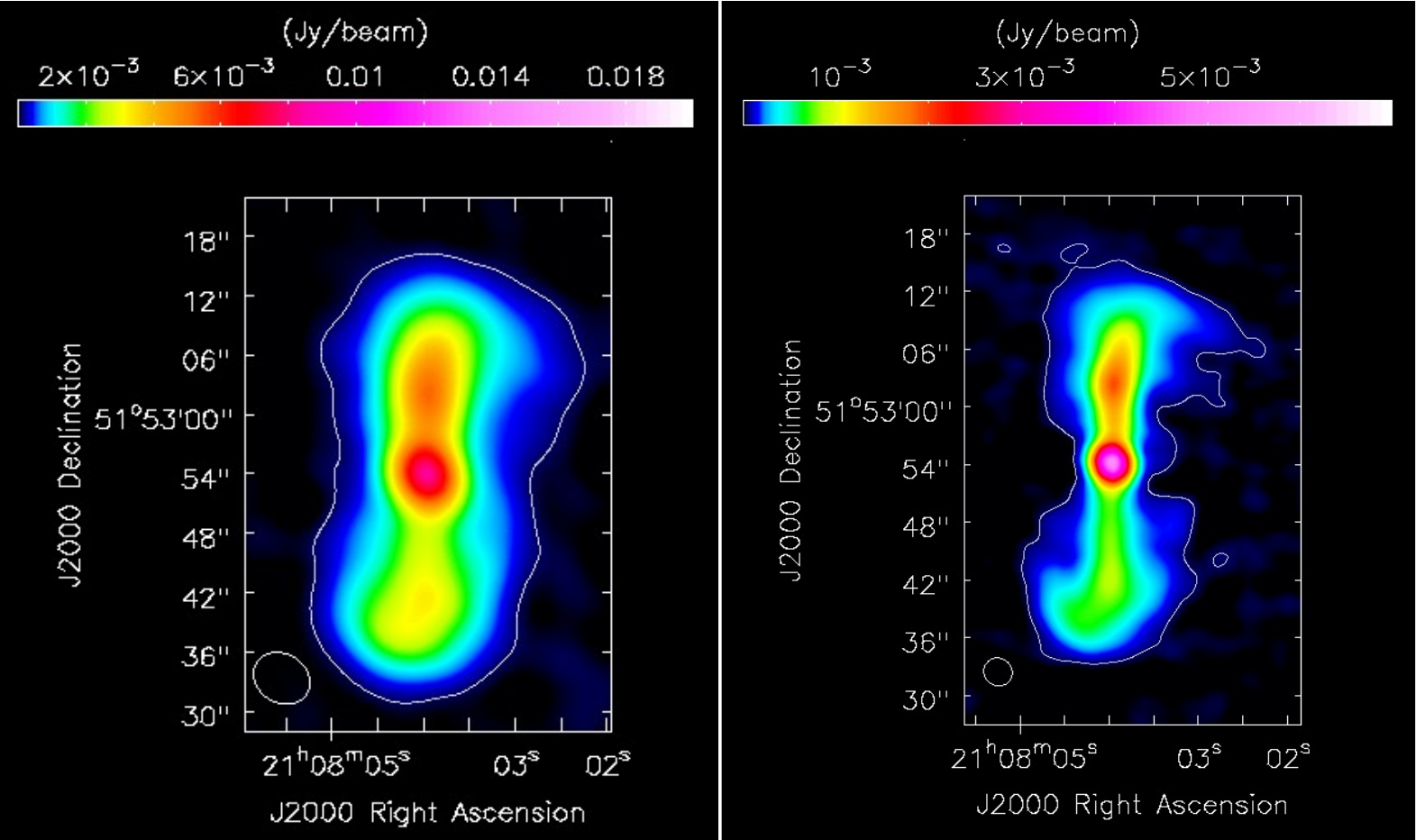}
    \caption{\textbf{Left:} Morphology of GMRT extended source (\gmrtext) in 400 MHz emission map obtained in uGMRT observation. The image is characterized by a synthesis beam $\sim 5.96'' \times 4.93''$ with position angle (P.A.) $=57.9^\circ$, shown in the bottom-left corner of the image. The average rms noise of the image is $\sim 23 \; \rm{\mu Jy}$. The contour shows the sources with 5$\sigma$ significance in the emission map. \textbf{Right:} Morphology of \gmrtext\ in 1.2 GHz emission map obtained in uGMRT observation. The image is characterized by a synthesis beam $\sim 3.02'' \times 2.86''$ with position angle (P.A.) $=52.5^\circ$ shown in the bottom-left corner of the image. The average rms noise of the image is $\sim 8 \; \rm{\mu Jy}$. Contore shows the sources with 5$\sigma$ significance in the emission map.}
    \label{fig:band3_5_img}
\end{figure}

The 400 MHz (band 3) and 1.2 GHz (band 5) images of \gmrtext\ are shown in Figure~ \ref{fig:band3_5_img}. The band 3 image has a synthesized beam of $5.96'' \times 4.93''$ (P.A. $=57.9^\circ$) with an rms noise of $\sim$23 $\mu$Jy beam$^{-1}$, while the band 5 image achieves a resolution of $3.02'' \times 2.86''$ (P.A. $=52.5^\circ$) with an rms of $\sim$8 $\mu$Jy beam$^{-1}$. The source morphology is consistent across all three bands (bands 3, 4, and 5), confirming the robustness of the double-lobed structure.

The source position in each band was determined by fitting a two-dimensional Gaussian within the $5\sigma$ emission contour; the fitted peak was adopted as the centroid position. The derived positions are summarized in Table~\ref{Tab:Position_and_flux}. Integrated flux densities were measured over the $5\sigma$ region, and their uncertainties were estimated following \citep{Rajpurohit_2022}:
\begin{equation}
\Delta S = \sqrt{(f S)^2 + N_{\mathrm{beam}},\sigma_{\mathrm{rms}}^2},
\end{equation}
where $S$ is the integrated flux density, $N_{\mathrm{beam}}$ is the number of synthesized beams within the source region, $\sigma_{\mathrm{rms}}$ is the image noise, and $f$ represents the fractional flux calibration uncertainty. We assume $f \; \sim $0.1 for uGMRT band 3 \citep{Chandra_2004}, 0.05 for band 4 and 0.025 for band 5 data \citep{Rajpurohit_2022}. The measured flux densities are listed in Table~\ref{Tab:Position_and_flux}.

\begin{figure}
   \centering
   \includegraphics[width=\linewidth]{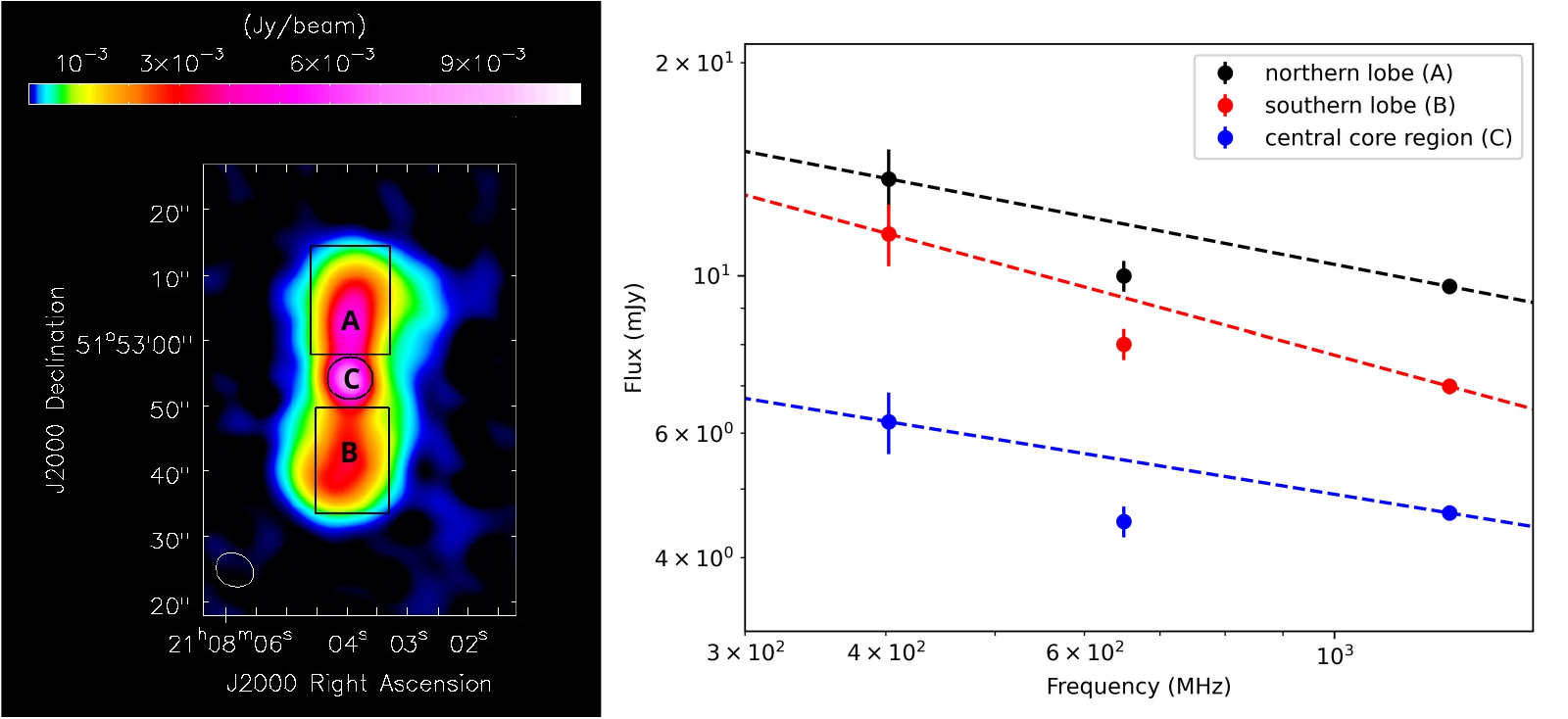}
  \caption{\textbf{Left:} uGMRT image of \gmrtext\ at 400 MHz, showing the regions used for spatially resolved spectral analysis. The rectangular boxes mark the northern (A) and southern (B) lobes, while the central circular region (C) denotes the core. The synthesized beam is shown in the bottom-left corner. \textbf{Right:} Spatially resolved radio spectra of the three regions, derived from flux densities measured in the resolution-matched images across bands 3, 4, and 5. The dashed lines represent power-law fits to the data points.
}
   \label{fig:spatial_variation}
\end{figure}

Apart from the integrated flux, in jeted source like \gmrtext, study of spatial variation of spectrum is also important, as it revals various physical properities such as the variation of magnetic field along the flow, variation of particle spectrum along the jet etc. We computed the spectrum of the source in three different region, as shown in the left panel of Fig \ref{fig:spatial_variation}. To investigate spatial variations in the spectrum, we first convolved the band 4 and band 5 images to match the angular resolution of band 3. Three regions were then defined within the source, as shown in Figure~\ref{fig:spatial_variation}: the northern lobe (A), the southern lobe (B), and the central core region (C). Flux densities at 400 MHz, 650 MHz and 1.2 GHz were extracted from identical regions in the resolution-matched images across bands 3, 4, and 5. These measurements were used to construct the spectra of regions A, B and C as shown in right panel of Fig~\ref{fig:spatial_variation}.

The combined three-band spectra could not be adequately described by a single power law. This could be due to the variability, as the band 4 observations were obtained two years earlier. Therefore, considering only the band 3 and band 5 data for the spectral slope estimation, it was found that the slope were -0.3, -0.4, and -0.26 for the lobe A, lobe B and core C respectively. This implies that the average photon spectra in region A, B, and C are produced by non-thermal electron populations with spectral indices -1.6, -1.8, and -1.52 respectively. The measured 650 MHz flux density deviates significantly from the value expected from the band 3 and band 5 spectrum. The absolute fractional deviation of the observed band 4 flux density from the expected value is 3.6$\sigma$, 3.06$\sigma$, and 4.4$\sigma$ for the northern lobe, southern lobe, and core, respectively. Since the band 3 and band 5 observations are nearly contemporaneous, whereas the band 4 observation was obtained approximately two years earlier, this discrepancy suggests intrinsic flux variability on a timescale of years. Such variability implies a characteristic source size of $\lesssim 1$ pc. We have also produced an index map using non-contemporaneous band 3 and band 5 data as shown in Fig. 3.

Following \cite{Urry_1995}, the jet inclination angle of \gmrtext\ is established using following equation,
\begin{equation}
\beta \cos \theta = \frac{J^{1/p} - 1}{J^{1/p} + 1},
\end{equation}
In this context, $J$ represents the ratio of jet to counterjet flux density, and $p$ is defined as 2+$\alpha$, assuming a continuous jet. Here, $\alpha$ denotes the spectral index of the source's integrated spectrum(Appendix - B, \cite{Urry_1995}). Using the integrated flux value, listed in table ~\ref{Tab:Position_and_flux}, the spectral index $\alpha$ is established to be 0.7. We are assuming relativistic flow, hence $\beta \sim 1$. Using these value the jet inclination angle of the source is established as $86.57^{\circ} \pm 1.15^{\circ}$.

\par  However, recent infrared observations \citep{Marti_2026_IR} identify a faint counterpart with $K_s \approx 19$, which disfavors a high-mass X-ray binary (HMXB) nature of the source, but the low mass X-ray binary (LMXB) scenario is still possible. LMXB with such high magnitude (with $K_s = 20.3$) is already reported in literature \citep{Eikenberry_2001}. The IR survey conducted by \cite{Marti_2026_IR} could not reveal any signature of SNR, PWN or even a shock wind environment like knot. Absence of any other accelerator strongly indicates that the dual lobed source is possibly producing PeV particles. In this work we explore the possibility of \gmrtext\ to explain the UHE emission in the framework of microquasar.

\begin{figure}
    \centering
    \includegraphics[width=0.5\linewidth]{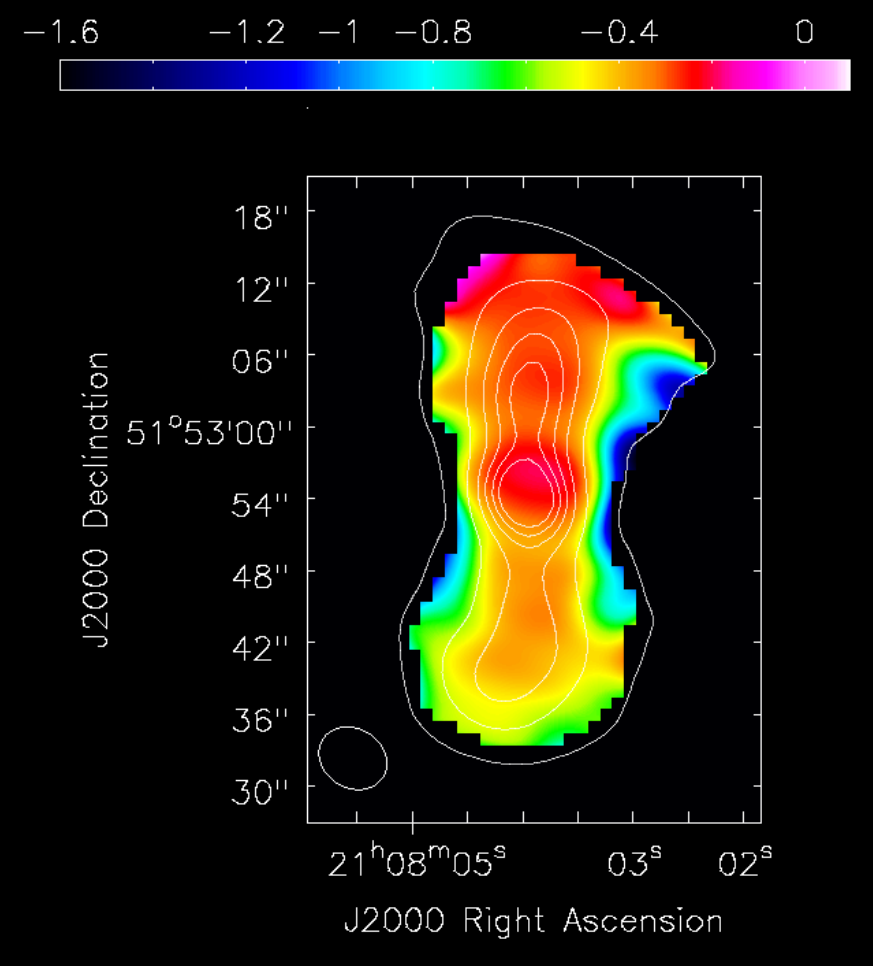}
    \caption{ Spectral index ($\alpha$) map of \gmrtext\ derived between uGMRT band 3 (400 MHz) and band 5 (1.2 GHz) after matching the angular resolution. The color scale represents the spatial variation of the spectral index across the source. Contours are overlaid from the band 5 (1.2 GHz) image at 10, 100, 200, 300, 400, and 500$\sigma$ levels.
}
    \label{fig:index_map}
\end{figure}

\begin{table*}[]
\caption{Measured positions (RA, Dec; J2000) and integrated flux densities of \gmrtext\ at different observing frequencies. Positions are derived from two-dimensional Gaussian fitting, and flux densities are measured within the $5\sigma$ emission region.}
\begin{center}
    \begin{tabular}{c c c c c c }
    \hline
    Observation band&Frequency&Observation Date&RA&Dec&Flux  \\
    &(MHz)& & (J2000)&(J2000)&(mJy)\\
    \hline
    \hline
		Band 3&402&07.06.2025 and 24.06.2025&21:08:03.958 $\pm$ 0.019 s&51.52.55.352 $\pm$ 1.048 arc sec&40.57 $\pm$ 4.05\\
    \hline
    Band 4&650&25.06.2023 and 11.07.2023&21:08:03.841 $\pm$ 0.017 s&51:52:56.680 $\pm$ 1.030 arc sec &27.6 $\pm$1.38\\
    \hline
    Band 5&1264&21.07.2025&21:08:03.954 $\pm$ 0.0094 s&51:52:56.411 $\pm$ 0.792 arc sec&26.1 $\pm$0.66\\
    \hline
    \end{tabular}
\label{Tab:Position_and_flux}
\end{center}
\end{table*}

\subsection{Multiwavelength spectra}

We constructed the multiwavelength spectral energy distribution (SED) using uGMRT data, X-ray upper limit and the VHE data points already reported in literature \citep{Adams_2025}. As the source was not resolved by LHAASO to core and jet regions, we took the integrated flux from the source for GMRT band 3, 4, and 5 to construct the SED. 
\par To begin with we considered the leptonic scenario to model the spectral energy distribution (SED) of the source. We assumed that the relativistic non-thermal electrons, present in the source, emit radiation by synchrotron process in the source magnetic field and by inverse Compton scattereing of 2.7~K cosmic microwave radiation. As jet inclination angle was large ($\sim 87^{\circ}$), therefore we did not consider any kind of Doppler boosting effect while evaluating the synchrotron and inverse Compton emissivities. The synchrotron and IC emissivities are calculated following \citet{Celli_2020} and \cite{Blumenthal_1970}, \cite{Lefa_2012} respectively. Following \citet{Adams_2025} we also assume that the non-thermal electrons follow the distribution
\begin{equation}
\frac{dN}{dE} = N_0
\left(\frac{E}{1 \,\rm{TeV}}\right)^{\alpha}
\exp\left[-\left(\frac{E}{E_{\rm cut}}\right)^\beta \right] \; \text{for all  } E \geq E_{\rm{min}}
\label{eq:ECPL}
\end{equation}
Where $N_0$ is the normalization constant, $\alpha$ is the spectral index, and $E_{\rm{cut}}$ is the cut-off energy of the spectrum. 

\par
\begin{figure}
    \centering
    \includegraphics[width=0.5\linewidth]{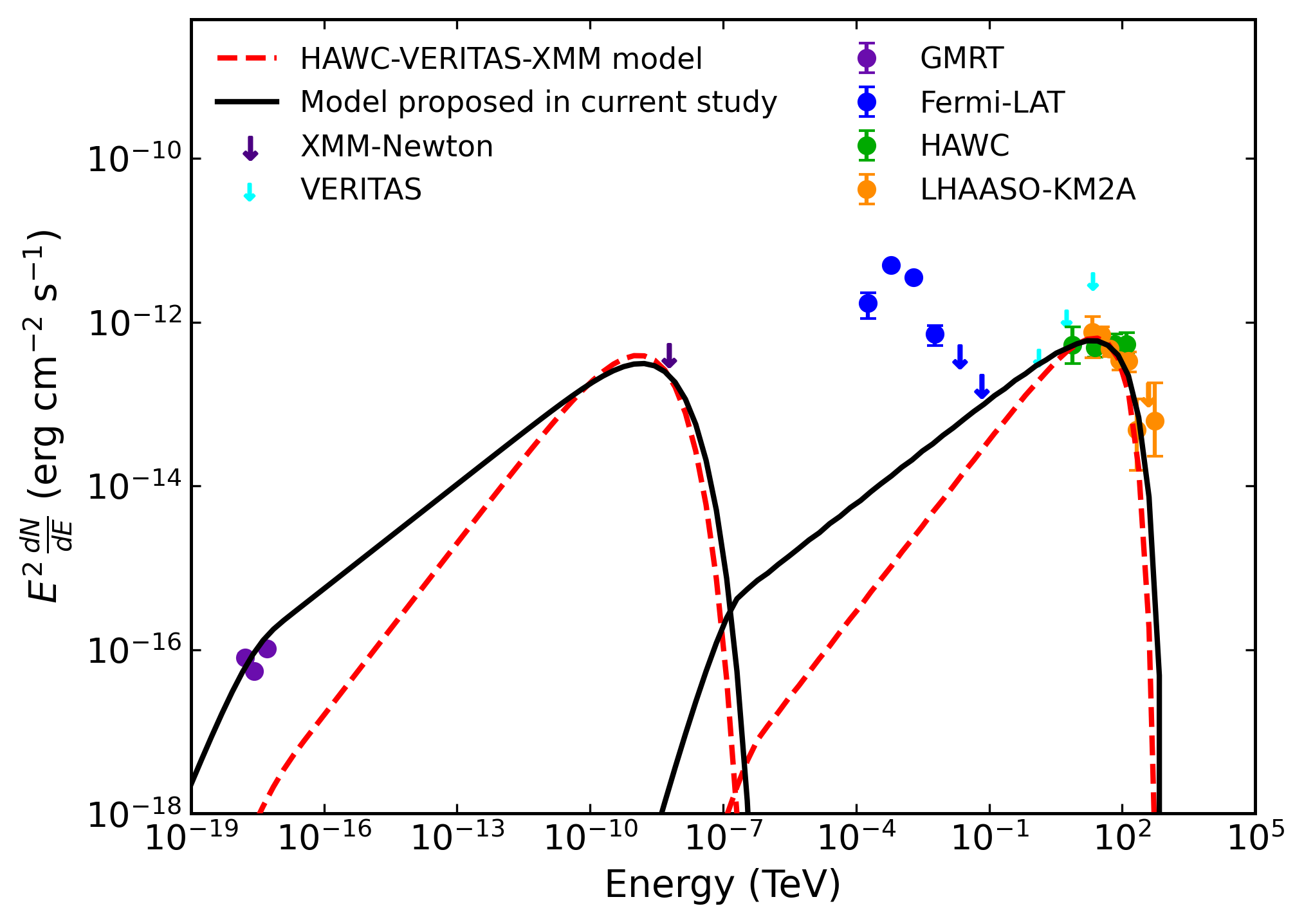}
    \caption{Broadband spectral energy distribution (SED) of LHAASO J2108+5157. The solid  lines represent the leptonic components with parameters derived in this work, while red dashed line shows the leptonic component corresponding to the parameter used in \cite{Adams_2025}. X-ray to ultra-high-energy (UHE) data points are taken from \cite{Adams_2025} and references therein, while the radio data points are derived in this work (band 3 and band 5) and Paper I (band 4).}
    \label{fig:sed_leptonic}
\end{figure}

 First, we use the model parameters $\alpha = 1.6$, $\beta = 2.0$, $E_{\rm{cut}} = 196$ TeV, and magnetic field $B \; = \; 10^{-6}$ G as reported by \citet{Adams_2025} in the present scenario. The model SED (red dashed line) along with the data are shown in Fig ~\ref{fig:sed_leptonic}. It is evident that the model cannot explain the radio data from uGMRT observation with the parameters used by \citet{Adams_2025}. To explain the radio observation in the leptonic scenario we need to take a steeper electron spectrum, but with a higher cut-off energy. We consider $\alpha = 2.15$ and $E_{\rm{cut}} = 300$ TeV.  A minimum electron energy $\sim 6 \times 10^{-3}$ TeV is necessary to model the radio data. Here magnetic field was kept fixed at $10^{-6}$ G. Thus the minimum electron of the electron distribution is constrained by the radio observation. The model curve is shown in Fig. ~\ref{fig:sed_leptonic} (black line). It is evident from the figure that the IC spectrum can explain the gamma ray spectrum upto $\sim$ 215~TeV energy. If we stretch the electron spectrum to higher energies by increasing the cut-off energy, it can explain the $\sim$ 537~TeV flux point. But the X-ray flux in 0.1--10~keV energy range violates the constrain set by XMM-Newton upper limit.  Considering the distance of the source to be $\sim$ 1 kpc and source size $\sim$~4.3~pc, the total energy content of the electron and magnetic field are found to be  $8.47\times 10^{45}$ erg and $3.11 \times 10^{45}$ erg respectively. It corresponds to an electron power $\sim$~10$^{37}$ erg/s.
 
\begin{figure}
    \centering
    \includegraphics[width=0.5\linewidth]{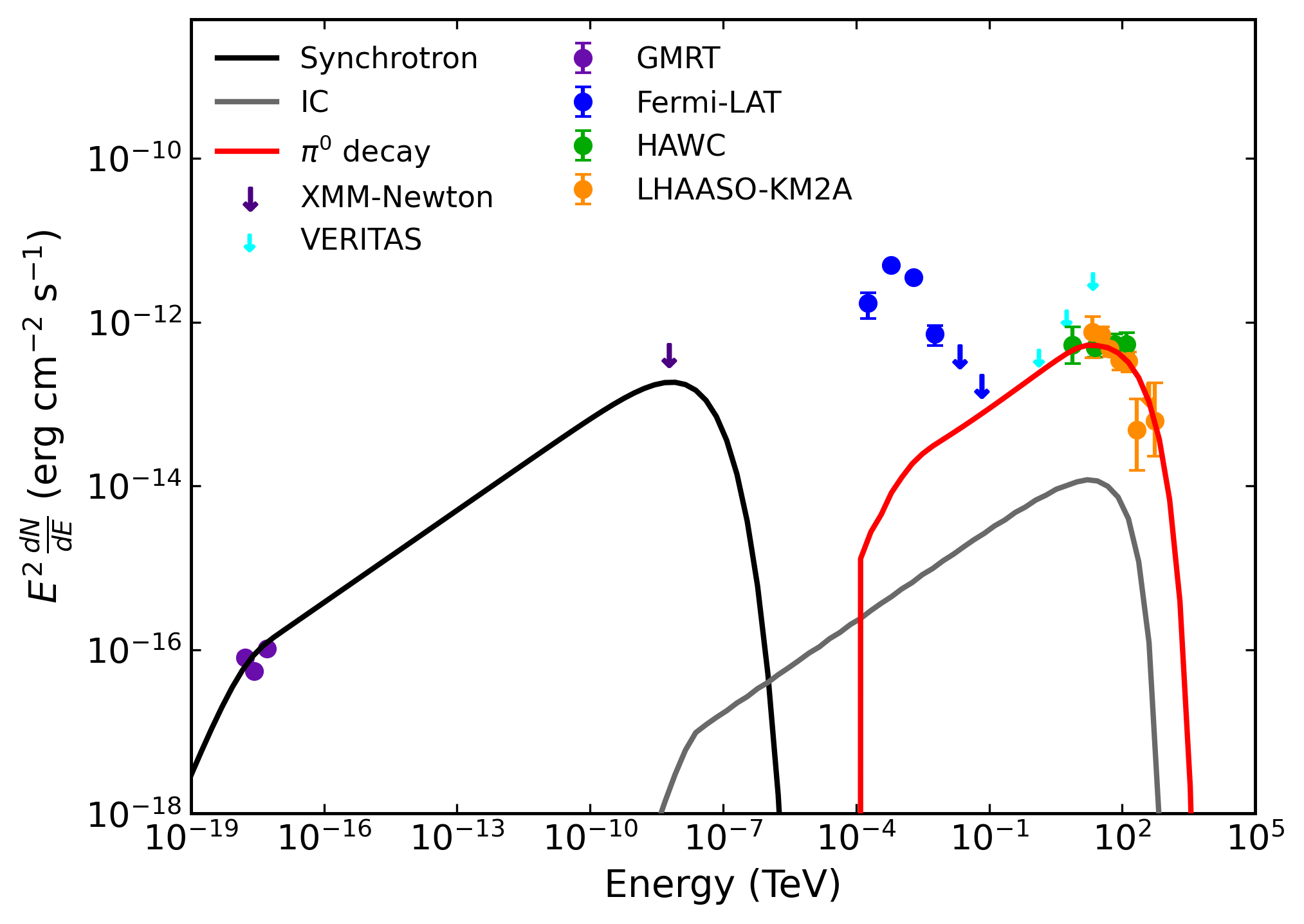}
    \caption{Broadband spectral energy distribution (SED) of LHAASO J2108+5157. The solid black and gray lines represent the leptonic components, corresponding to synchrotron emission and inverse Compton scattering of cosmic microwave background (CMB) photons, respectively. The red line shows the hadronic $\gamma$-ray emission produced via neutral pion ($\pi^{0}$) decay in the molecular cloud. X-ray to ultra-high-energy (UHE) data points are taken from \cite{Adams_2025} and references therein, while the radio data points are derived in this work (band 3 and band 5) and Paper I (band 4).}
    \label{fig:sed_hadronic}
\end{figure}

 \noindent As a possible alternative to the leptonic scenario we also tested a lepto-hadronic scenario to model the present SED of the source. We assumed that electrons and protons are accelerated to relativistic energies by the same acceleration process. Both electrons and protons have same spectral form as given by Eq. (\ref{eq:ECPL}), but with different spectral parameters. It is assumed that, the non-thermal relativistic protons interact with the molecular cloud around the source to produce $\pi^0$ which in turn decay into gamma-rays \citep{Abe_2023,Fuente_2023a,Fuente_2023b}.  The $\gamma$-ray emissivity from $\pi^0$ decay is calculated following \cite{Celli_2020}. Contribution of non-thermal electrons were estimated in the same way as it was done in case of  leptonic model. The leptonic component models the radio - to - X-ray part of the SED while the hadronic component dominates the gamma-ray part of the SED. The model spectrum along with the data points are shown in Fig. ~\ref{fig:sed_hadronic}. The resultant SED follows the constraint set by the XMM-Newton observation as well as it can explain the $\sim$ 537~TeV flux point observed with KM2A. The parameter values used in the model for electron and proton spectra are given in Table ~\ref{tab:params}. It is important to note that the cut-off energy required for the proton $\sim$ 1800 TeV (1.8 PeV). This implies that the particle acceleration processes would have to be such that it produces PeV protons in the source. The energy content of electrons, protons, and magnetic field is estimated to be $4.24 \times 10^{44}$, $1.28 \times 10^{46}$, and $7.78 \times 10^{46}$ erg, respectively. The estimated power in electrons and protons are found to be $\sim$~10$^{36}$ erg/s and 10$^{38}$ erg/s respectively.

 \noindent Thus, through spectral modeling we showed that both leptonic and leptohadronic scenarios can be used to understand the SED. The source magnetic field, electron and proton power obtained in our studies are quite consistent with those obtained through detail modeling of microquasars already published in literature, for example, \citet{Abaroa_2024} for microquasar S26.  As we did not perform any $\chi^2$-fitting of the data with the models, we cannot statistically rule out any particular model. But the leptohadronic scenario can reproduce the observed flux at the highest energy bin of KM2A detector, while this is not possible in leptonic scenario obeying the observed upper limit in X-ray energy. Therefore, deep observations with KM2A are necessary to have a better understanding of the source.

\begin{table*}[t]
	\centering
	\caption{Parameters used to fit the broadband spectral energy distributions of the PeVatron sources LHAASO J2108+5157 using Leptonic and Lepto-hadronic approach.}
	\begin{tabular}{llccc}
		\toprule
		& \textbf{Physical Parameter} & \textbf{Leptonic} & \textbf{Lepto-hadronic} \\
		\midrule
		\multirow{4}{*}{Electron Spectrum Parameter} 
		& Index, $\alpha$ & -2.15  & -2.25 \\
		& Cut-off Energy, $E_{cut}$ & 300 TeV  & 300 TeV \\
        & Injection Energy, $E_{min}$ & $6\times 10 ^{-3}$ TeV & $2\times 10 ^{-3}$ TeV \\
		& $\beta$ & 2  & 2\\
		\midrule
		\multirow{4}{*}{Proton Spectrum Parameter} 
		& Index, $\alpha$ & -- & -1.7 \\
		& Cut-off Energy, $E_{cut}$ & -- & 1.8 PeV \\
        & Injection Energy, $E_{min}$ & -- & $1 \times 10^{-2}$ TeV \\
		& $\beta$ & -- & 2\\
		\midrule
		\multirow{3}{*}{Cloud Property} 
		& Radius of the cloud, $R_{cl}$ & -- & 16.31 pc \\
		& number density of the cloud $n_H$ & -- & 100 $\rm{cm^{-3}}$\\
        & Distance of the source/cloud, $d$ & 1.0 kpc & 1.0 kpc \\
		\midrule
		\multirow{2}{*}{Source Propertity} 
		& Radius of the source $R_{sc} $& 4.3 pc& 4.3 pc\\
		& Magnetic Field, $B$ & 1 $\mu G$ & $5$ $\mu G$\\
		\bottomrule
	\end{tabular}
	
	\label{tab:params}
\end{table*}

\section{Conclusion}

We report deep uGMRT observations of the PeVatron candidate LHAASO J2108+5157, revealing a persistent double-lobed radio source (\gmrtext) with consistent morphology across bands 3, 4, and 5. This complets the spectral and morphological study of the possible candidate for the dark PeVatron LHAASO J2108+5157 detected by LHAASO. It was found that
\begin{itemize}
    \item the core and lobes show optically thin radio spectrum in frequency range 200 MHz - 1.4 GHz indicating the presence of relativistic non-thermal electrons in the source.
    \item the 650 MHz flux density shows significant deviation from the expected power law estimate at this frequency. The band 3 and 5 observations were near-simultaneous (separated by a month) while band 4 (650 MHz) observation was separated by approximately 2 years. Hence, the deviation in flux can be attributed the inherent flux variability. The presence of any such variability needs to be ascertained by future time-lapse observations.
    \item the spectral index map generated with band 3 and band 5 data shows the presence of a flowing radiative matter in the jet-like structure of the source.
    \item the observed spectral energy distribution were modeled by pure leptonic as well as lepto-hadronic models. The lepto-hadronic scenario models the SED better than the leptonic scenario. But
    deeper observation of the source with KM2A is necessary to constrain the model further.
\end{itemize}
Finally, although \citet{Marti_2026_IR} argued that the uGMRT source can not be a microquasar with high mass companion, the possibility that it is a mcroquasar with a low mass companion having IR
magnitude $K_s\,>\,19$ can not be ruled out. For example, the microquasar GRS~1758--258 has IR 
magnitude of $K_s\,=\,20.3$ \citep{Eikenberry_2001}. Therefore, more observations are required to 
identify the true nature of the mysterious PeVatron.

\begin{acknowledgments}
We sincerely thank the anonymous reviewer for their valuable suggestions, which have significantly improved the quality of the manuscript.
We thank the staff of GMRT that allowed these observations to be made. GMRT is run by National Centre for
Radio Astrophysics of the Tata Institute of fundamental research. S.R. acknowledges the support of
of the Department of Atomic Energy, under project no. 12-R\&D-TFR-5.02-0700. GKM acknowledges Yogesh Ramesh Pawar, HBNI, for his genuine help in downloading the bulk dataset.
\end{acknowledgments}





%
\facilities{GMRT}

\software{CASA}





\bibliography{sample701}{}

@ARTICLE{Mahanta_2025,
       author = {{Krishna Mahanta}, Gunindra and {Roy}, Subhashis and {Godambe}, Sagar and {Ghosal}, Bitan and {Bhatt}, Nilay and {Bhattacharyya}, Subir},
        title = "{Low frequency radio observation of the dark PeVatron 1LHAASO J2108+5153u using uGMRT}",
      journal = {Journal of High Energy Astrophysics},
         year = 2025,
        month = jul,
       volume = {47},
          eid = {100381},
        pages = {100381},
          doi = {10.1016/j.jheap.2025.100381},
       adsurl = {https://ui.adsabs.harvard.edu/abs/2025JHEAp..4700381K}
}

@ARTICLE{Butt_2008,
       author = {{Butt}, Y.~M. and {Combi}, J.~A. and {Drake}, J. and {Finley}, J.~P. and {Konopelko}, A. and {Lister}, M. and {Rodriguez}, J. and {Shepherd}, D.},
        title = "{TeV J2032+4130: a not-so-dark accelerator?}",
      journal = {\mnras},
         year = 2008,
        month = apr,
       volume = {385},
       number = {4},
        pages = {1764-1770},
          doi = {10.1111/j.1365-2966.2008.12959.x},
archivePrefix = {arXiv},
       eprint = {astro-ph/0611731},
 primaryClass = {astro-ph},
       adsurl = {https://ui.adsabs.harvard.edu/abs/2008MNRAS.385.1764B}
}

@ARTICLE{Marti_2026,
       author = {{Mart{\'\i}}, Josep and {Luque-Escamilla}, Pedro L.},
        title = "{Alignment of radio jets in the microquasar V4641 Sagittarii with its high-energy structures}",
      journal = {\mnras},
         year = 2026,
        month = jan,
       volume = {545},
       number = {3},
          eid = {staf2104},
        pages = {staf2104},
          doi = {10.1093/mnras/staf2104},
archivePrefix = {arXiv},
       eprint = {2511.19695},
 primaryClass = {astro-ph.HE},
       adsurl = {https://ui.adsabs.harvard.edu/abs/2026MNRAS.545f2104M}
}

@ARTICLE{Elder_2026,
       author = {{Edler}, H.~W. and {Arias}, M. and {Botteon}, A. and {Bassa}, C.~G.},
        title = "{Multi-wavelength insights into the pulsar wind nebula candidate near 1LHAASO J0343+5254u: an obscured merging galaxy cluster?}",
      journal = {arXiv e-prints},
         year = 2026,
        month = mar,
          eid = {arXiv:2603.03164},
        pages = {arXiv:2603.03164},
          doi = {10.48550/arXiv.2603.03164},
archivePrefix = {arXiv},
       eprint = {2603.03164},
 primaryClass = {astro-ph.HE},
       adsurl = {https://ui.adsabs.harvard.edu/abs/2026arXiv260303164E}
}

@ARTICLE{Rajpurohit_2022,
       author = {{Rajpurohit}, K. and {van Weeren}, R.~J. and {Hoeft}, M. and {Vazza}, F. and {Brienza}, M. and {Forman}, W. and {Wittor}, D. and {Dom{\'\i}nguez-Fern{\'a}ndez}, P. and {Rajpurohit}, S. and {Riseley}, C.~J. and {Botteon}, A. and {Osinga}, E. and {Brunetti}, G. and {Bonnassieux}, E. and {Bonafede}, A. and {Rajpurohit}, A.~S. and {Stuardi}, C. and {Drabent}, A. and {Br{\"u}ggen}, M. and {Dallacasa}, D. and {Shimwell}, T.~W. and {R{\"o}ttgering}, H.~J.~A. and {de Gasperin}, F. and {Miley}, G.~K. and {Rossetti}, M.},
        title = "{Deep Low-frequency Radio Observations of A2256. I. The Filamentary Radio Relic}",
      journal = {\apj},
         year = 2022,
        month = mar,
       volume = {927},
       number = {1},
          eid = {80},
        pages = {80},
          doi = {10.3847/1538-4357/ac4708},
archivePrefix = {arXiv},
       eprint = {2111.04449},
 primaryClass = {astro-ph.CO},
       adsurl = {https://ui.adsabs.harvard.edu/abs/2022ApJ...927...80R}
}

@ARTICLE{Chandra_2004,
       author = {{Chandra}, P. and {Ray}, A. and {Bhatnagar}, S.},
        title = "{The Late-Time Radio Emission from SN 1993J at Meter Wavelengths}",
      journal = {\apj},
         year = 2004,
        month = sep,
       volume = {612},
       number = {2},
        pages = {974-987},
          doi = {10.1086/422675},
archivePrefix = {arXiv},
       eprint = {astro-ph/0405448},
 primaryClass = {astro-ph},
       adsurl = {https://ui.adsabs.harvard.edu/abs/2004ApJ...612..974C}
}

@ARTICLE{Cao_2021,
       author = {{Cao}, Zhen and {Aharonian}, F.~A. and {An}, Q. and {Axikegu}, L.~X., Bai and {Bai}, Y.~X. and {Bao}, Y.~W. and {Bastieri}, D. and {Bi}, X.~J. and {Bi}, Y.~J. and {Cai}, H. and {Cai}, J.~T. and {Cao}, Zhe and {Chang}, J. and {Chang}, J.~F. and {Chang}, X.~C. and {Chen}, B.~M. and {Chen}, J. and {Chen}, L. and {Chen}, Liang and {Chen}, Long and {Chen}, M.~J. and {Chen}, M.~L. and {Chen}, Q.~H. and {Chen}, S.~H. and {Chen}, S.~Z. and {Chen}, T.~L. and {Chen}, X.~L. and {Chen}, Y. and {Cheng}, N. and {Cheng}, Y.~D. and {Cui}, S.~W. and {Cui}, X.~H. and {Cui}, Y.~D. and {Dai}, B.~Z. and {Dai}, H.~L. and {Dai}, Z.~G. and {Danzengluobu} and {della Volpe}, D. and {D'Ettorre Piazzoli}, B. and {Dong}, X.~J. and {Fan}, J.~H. and {Fan}, Y.~Z. and {Fan}, Z.~X. and {Fang}, J. and {Fang}, K. and {Feng}, C.~F. and {Feng}, L. and {Feng}, S.~H. and {Feng}, Y.~L. and {Gao}, B. and {Gao}, C.~D. and {Gao}, Q. and {Gao}, W. and {Ge}, M.~M. and {Geng}, L.~S. and {Gong}, G.~H. and {Gou}, Q.~B. and {Gu}, M.~H. and {Guo}, J.~G. and {Guo}, X.~L. and {Guo}, Y.~Q. and {Guo}, Y.~Y. and {Han}, Y.~A. and {He}, H.~H. and {He}, H.~N. and {He}, J.~C. and {He}, S.~L. and {He}, X.~B. and {He}, Y. and {Heller}, M. and {Hor}, Y.~K. and {Hou}, C. and {Hou}, X. and {Hu}, H.~B. and {Hu}, S. and {Hu}, S.~C. and {Hu}, X.~J. and {Huang}, D.~H. and {Huang}, Q.~L. and {Huang}, W.~H. and {Huang}, X.~T. and {Huang}, Z.~C. and {Ji}, F. and {Ji}, X.~L. and {Jia}, H.~Y. and {Jiang}, K. and {Jiang}, Z.~J. and {Jin}, C. and {Kuleshov}, D. and {Levochkin}, K. and {Li}, B.~B. and {Li}, Cong and {Li}, Cheng and {Li}, F. and {Li}, H.~B. and {Li}, H.~C. and {Li}, H.~Y. and {Li}, J. and {Li}, K. and {Li}, W.~L. and {Li}, X. and {Li}, Xin and {Li}, X.~R. and {Li}, Y. and {Li}, Y.~Z. and {Li}, Zhe and {Li}, Zhuo and {Liang}, E.~W. and {Liang}, Y.~F. and {Lin}, S.~J. and {Liu}, B. and {Liu}, C. and {Liu}, D. and {Liu}, H. and {Liu}, H.~D. and {Liu}, J. and {Liu}, J.~L. and {Liu}, J.~S. and {Liu}, J.~Y. and {Liu}, M.~Y. and {Liu}, R.~Y. and {Liu}, S.~M. and {Liu}, W. and {Liu}, Y.~N. and {Liu}, Z.~X. and {Long}, W.~J. and {Lu}, R. and {Lv}, H.~K. and {Ma}, B.~Q. and {Ma}, L.~L. and {Ma}, X.~H. and {Mao}, J.~R. and {Masood}, A. and {Mitthumsiri}, W. and {Montaruli}, T. and {Nan}, Y.~C. and {Pang}, B.~Y. and {Pattarakijwanich}, P. and {Pei}, Z.~Y. and {Qi}, M.~Y. and {Ruffolo}, D. and {Rulev}, V. and {S{\'a}iz}, A. and {Shao}, L. and {Shchegolev}, O. and {Sheng}, X.~D. and {Shi}, J.~R. and {Song}, H.~C. and {Stenkin}, Yu. V. and {Stepanov}, V. and {Sun}, Q.~N. and {Sun}, X.~N. and {Sun}, Z.~B. and {Tam}, P.~H.~T. and {Tang}, Z.~B. and {Tian}, W.~W. and {Wang}, B.~D. and {Wang}, C. and {Wang}, H. and {Wang}, H.~G. and {Wang}, J.~C. and {Wang}, J.~S. and {Wang}, L.~P. and {Wang}, L.~Y. and {Wang}, R.~N. and {Wang}, W. and {Wang}, W. and {Wang}, X.~G. and {Wang}, X.~J. and {Wang}, X.~Y. and {Wang}, Y.~D. and {Wang}, Y.~J. and {Wang}, Y.~P. and {Wang}, Zheng and {Wang}, Zhen and {Wang}, Z.~H. and {Wang}, Z.~X. and {Wei}, D.~M. and {Wei}, J.~J. and {Wei}, Y.~J. and {Wen}, T. and {Wu}, C.~Y. and {Wu}, H.~R. and {Wu}, S. and {Wu}, W.~X. and {Wu}, X.~F. and {Xi}, S.~Q. and {Xia}, J. and {Xia}, J.~J. and {Xiang}, G.~M. and {Xiao}, G. and {Xiao}, H.~B. and {Xin}, G.~G. and {Xin}, Y.~L. and {Xing}, Y. and {Xu}, D.~L. and {Xu}, R.~X. and {Xue}, L. and {Yan}, D.~H. and {Yang}, C.~W.},
        title = "{Ultrahigh-energy photons up to 1.4 petaelectronvolts from 12 {\ensuremath{\gamma}}-ray Galactic sources}",
      journal = {\nat},
         year = 2021,
        month = jun,
       volume = {594},
       number = {7861},
        pages = {33-36},
          doi = {10.1038/s41586-021-03498-z},
       adsurl = {https://ui.adsabs.harvard.edu/abs/2021Natur.594...33C}
}

@article{Adams_2025,
doi = {10.3847/1538-4357/adf439},
url = {https://doi.org/10.3847/1538-4357/adf439},
year = {2025},
month = {sep},
publisher = {The American Astronomical Society},
volume = {991},
number = {2},
pages = {192},
author = {Adams, C. B. and Bangale, P. and Benbow, W. and Buckley, J. H. and Chen, Y. and Christiansen, J. L. and Chromey, A. J. and Escobar Godoy, M. and Feldman, S. and Feng, Q. and Foote, J. and Fortson, L. and Furniss, A. and Hanlon, W. and Hervet, O. and Hinrichs, C. E. and Holder, J. and Hughes, Z. and Humensky, T. B. and Jin, W. and Kaaret, P. and Kertzman, M. and Kherlakian, M. and Kieda, D. and Kleiner, T. K. and Korzoun, N. and Kumar, S. and Lang, M. J. and Lundy, M. and Maier, G. and Millard, M. J. and Moriarty, P. and Mukherjee, R. and Ning, W. and Ong, R. A. and Pohl, M. and Pueschel, E. and Quinn, J. and Rabinowitz, P. L. and Ragan, K. and Reynolds, P. T. and Ribeiro, D. and Roache, E. and Sadeh, I. and Saha, L. and Sembroski, G. H. and Shang, R. and Tak, D. and Talluri, A. K. and Tucci, J. V. and Valverde, J. and Williams, D. A. and Wong, S. L. and Woo, J. and (The VERITAS collaboration) and Alfaro, R. and Alvarez, C. and Arteaga-Velázquez, J. C. and Avila Rojas, D. and Babu, R. and Belmont-Moreno, E. and Bernal, A. and Caballero-Mora, K. S. and Carramiñana, A. and Casanova, S. and Cotti, U. and Cotzomi, J. and De la Fuente, E. and de León, C. and Depaoli, D. and Desiati, P. and Di Lalla, N. and Diaz Hernandez, R. and DuVernois, M. A. and Engel, K. and Ergin, T. and Espinoza, C. and Fan, K. L. and Fraija, N. and Fraija, S. and García-González, J.A. and Garfias, F. and Gonzalez Muñoz, A. and González, M. M. and Goodman, J. A. and Groetsch, S. and Harding, J. P. and Hernández-Cadena, S. and Herzog, I. and Huang, D. and Hueyotl-Zahuantitla, F. and Hüntemeyer, P. and Iriarte, A. and Kaufmann, S. and Lara, A. and Lee, J. and León Vargas, H. and Longinotti, A. L. and Luis-Raya, G. and Malone, K. and Martinez, O. and Martínez-Castro, J. and Matthews, J. A. and Miranda-Romagnoli, P. and Morales-Soto, J. A. and Moreno, E. and Araya, M. and Mostafá, M. and Najafi, M. and Nayerhoda, A. and Nellen, L. and Omodei, N. and Ponce, E. and Pérez-Pérez, E. G. and Rho, C. D. and Rosa-González, D. and Roth, M. and Salazar, H. and Sandoval, A. and Schneider, M. and Serna-Franco, J. and Smith, A. J. and Son, Y. and Springer, R. W. and Tibolla, O. and Tollefson, K. and Torres, I. and Torres-Escobedo, R. and Turner, R. and Ureña-Mena, F. and Varela, E. and Villaseñor, L. and Wang, X. and Wang, Z. and Watson, I. J. and Wu, H. and Yu, S. and Yun-Cárcamo, S. and Zhou, H. and Martin, M. and (The HAWC collaboration) and Mori, Kaya and Hailey, Charles J. and Safi-Harb, Samar and Zhang, Shuo and (The XMM-Newton collaboration)},
title = {HAWC, VERITAS, Fermi-LAT, and XMM-Newton Follow-up Observations of the Unidentified Ultra-high-energy Gamma-Ray Source LHAASO J2108+5157},
journal = {The Astrophysical Journal},
}

@ARTICLE{Cao_discovery_2021,
       author = {{Cao}, Zhen and {Aharonian}, F. and {An}, Q. and {Axikegu} and {Bai}, L.~X. and {Bai}, Y.~X. and {Bao}, Y.~W. and {Bastieri}, D. and {Bi}, X.~J. and {Bi}, Y.~J. and {Cai}, H. and {Cai}, J.~T. and {Cao}, Zhe and {Chang}, J. and {Chang}, J.~F. and {Chen}, B.~M. and {Chen}, E.~S. and {Chen}, J. and {Chen}, Liang and {Chen}, Long and {Chen}, M.~J. and {Chen}, M.~L. and {Chen}, Q.~H. and {Chen}, S.~H. and {Chen}, S.~Z. and {Chen}, T.~L. and {Chen}, X.~L. and {Chen}, Y. and {Cheng}, N. and {Cheng}, Y.~D. and {Cui}, S.~W. and {Cui}, X.~H. and {Cui}, Y.~D. and {Piazzoli}, B. D'Ettorre and {Dai}, B.~Z. and {Dai}, H.~L. and {Dai}, Z.~G. and {Dan-Zeng-Luo-Bu}, Dan-Zeng and {Volpe}, D. della and {Dong}, X.~J. and {Duan}, K.~K. and {Fan}, J.~H. and {Fan}, Y.~Z. and {Fan}, Z.~X. and {Fang}, J. and {Fang}, K. and {Feng}, C.~F. and {Feng}, L. and {Feng}, S.~H. and {Feng}, Y.~L. and {Gao}, B. and {Gao}, C.~D. and {Gao}, L.~Q. and {Gao}, Q. and {Gao}, W. and {Ge}, M.~M. and {Geng}, L.~S. and {Gong}, G.~H. and {Gou}, Q.~B. and {Gu}, M.~H. and {Guo}, F.~L. and {Guo}, J.~G. and {Guo}, X.~L. and {Guo}, Y.~Q. and {Guo}, Y.~Y. and {Han}, Y.~A. and {He}, H.~H. and {He}, H.~N. and {He}, J.~C. and {He}, S.~L. and {He}, X.~B. and {He}, Y. and {Heller}, M. and {Hor}, Y.~K. and {Hou}, C. and {Hu}, H.~B. and {Hu}, S. and {Hu}, S.~C. and {Hu}, X.~J. and {Huang}, D.~H. and {Huang}, Q.~L. and {Huang}, W.~H. and {Huang}, X.~T. and {Huang}, X.~Y. and {Huang}, Z.~C. and {Ji}, F. and {Ji}, X.~L. and {Jia}, H.~Y. and {Jiang}, K. and {Jiang}, Z.~J. and {Jin}, C. and {Ke}, T. and {Kuleshov}, D. and {Levochkin}, K. and {Li}, B.~B. and {Li}, Cheng and {Li}, Cong and {Li}, F. and {Li}, H.~B. and {Li}, H.~C. and {Li}, H.~Y. and {Li}, J. and {Li}, K. and {Li}, W.~L. and {Li}, X.~R. and {Li}, Xin and {Li}, Xin and {Li}, Y. and {Li}, Y.~Z. and {Li}, Zhe and {Li}, Zhuo and {Liang}, E.~W. and {Liang}, Y.~F. and {Lin}, S.~J. and {Liu}, B. and {Liu}, C. and {Liu}, D. and {Liu}, H. and {Liu}, H.~D. and {Liu}, J. and {Liu}, J.~L. and {Liu}, J.~S. and {Liu}, J.~Y. and {Liu}, M.~Y. and {Liu}, R.~Y. and {Liu}, S.~M. and {Liu}, W. and {Liu}, Y. and {Liu}, Y.~N. and {Liu}, Z.~X. and {Long}, W.~J. and {Lu}, R. and {Lv}, H.~K. and {Ma}, B.~Q. and {Ma}, L.~L. and {Ma}, X.~H. and {Mao}, J.~R. and {Masood}, A. and {Min}, Z. and {Mitthumsiri}, W. and {Montaruli}, T. and {Nan}, Y.~C. and {Pang}, B.~Y. and {Pattarakijwanich}, P. and {Pei}, Z.~Y. and {Qi}, M.~Y. and {Qi}, Y.~Q. and {Qiao}, B.~Q. and {Qin}, J.~J. and {Ruffolo}, D. and {Rulev}, V. and {S{\'a}iz}, A. and {Shao}, L. and {Shchegolev}, O. and {Sheng}, X.~D. and {Shi}, J.~Y. and {Song}, H.~C. and {Stenkin}, Yu. V. and {Stepanov}, V. and {Su}, Y. and {Sun}, Q.~N. and {Sun}, X.~N. and {Sun}, Z.~B. and {Tam}, P.~H.~T. and {Tang}, Z.~B. and {Tian}, W.~W. and {Wang}, B.~D. and {Wang}, C. and {Wang}, H. and {Wang}, H.~G. and {Wang}, J.~C. and {Wang}, J.~S. and {Wang}, L.~P. and {Wang}, L.~Y. and {Wang}, R.~N. and {Wang}, W. and {Wang}, W. and {Wang}, X.~G. and {Wang}, X.~J. and {Wang}, X.~Y. and {Wang}, Y. and {Wang}, Y.~D. and {Wang}, Y.~J. and {Wang}, Y.~P. and {Wang}, Z.~H. and {Wang}, Z.~X. and {Wang}, Zhen and {Wang}, Zheng and {Wei}, D.~M. and {Wei}, J.~J. and {Wei}, Y.~J. and {Wen}, T. and {Wu}, C.~Y. and {Wu}, H.~R. and {Wu}, S. and {Wu}, W.~X. and {Wu}, X.~F. and {Xi}, S.~Q. and {Xia}, J. and {Xia}, J.~J. and {Xiang}, G.~M. and {Xiao}, D.~X. and {Xiao}, G. and {Xiao}, H.~B. and {Xin}, G.~G. and {Xin}, Y.~L. and {Xing}, Y. and {Xu}, D.~L. and {Xu}, R.~X. and {Xue}, L. and {Yan}, D.~H. and {Yan}, J.~Z. and {Yang}, C.~W. and {Yang}, F.~F. and {Yang}, J.~Y. and {Yang}, L.~L. and {Yang}, M.~J. and {Yang}, R.~Z. and {Yang}, S.~B. and {Yao}, Y.~H. and {Yao}, Z.~G. and {Ye}, Y.~M. and {Yin}, L.~Q. and {Yin}, N. and {You}, X.~H. and {You}, Z.~Y. and {Yu}, Y.~H. and {Yuan}, Q. and {Zeng}, H.~D. and {Zeng}, T.~X. and {Zeng}, W. and {Zeng}, Z.~K. and {Zha}, M. and {Zhai}, X.~X. and {Zhang}, B.~B. and {Zhang}, H.~M. and {Zhang}, H.~Y. and {Zhang}, J.~L. and {Zhang}, J.~W. and {Zhang}, L.~X. and {Zhang}, Li and {Zhang}, Lu and {Zhang}, P.~F. and {Zhang}, P.~P. and {Zhang}, R. and {Zhang}, S.~R. and {Zhang}, S.~S. and {Zhang}, X. and {Zhang}, X.~P. and {Zhang}, Y.~F. and {Zhang}, Y.~L. and {Zhang}, Yi and {Zhang}, Yong and {Zhao}, B. and {Zhao}, J. and {Zhao}, L. and {Zhao}, L.~Z. and {Zhao}, S.~P. and {Zheng}, F. and {Zheng}, Y. and {Zhou}, B. and {Zhou}, H. and {Zhou}, J.~N. and {Zhou}, P. and {Zhou}, R. and {Zhou}, X.~X. and {Zhu}, C.~G. and {Zhu}, F.~R. and {Zhu}, H. and {Zhu}, K.~J. and {Zuo}, X.},
        title = "{Discovery of the Ultrahigh-energy Gamma-Ray Source LHAASO J2108+5157}",
      journal = {\apjl},
         year = 2021,
        month = oct,
       volume = {919},
       number = {2},
          eid = {L22},
        pages = {L22},
          doi = {10.3847/2041-8213/ac2579},
archivePrefix = {arXiv},
       eprint = {2106.09865},
 primaryClass = {astro-ph.HE},
       adsurl = {https://ui.adsabs.harvard.edu/abs/2021ApJ...919L..22C}
}

@ARTICLE{Abe_2023,
       author = {{Abe}, S. and {Aguasca-Cabot}, A. and {Agudo}, I. and {Alvarez Crespo}, N. and {Antonelli}, L.~A. and {Aramo}, C. and {Arbet-Engels}, A. and {Artero}, M. and {Asano}, K. and {Aubert}, P. and {Baktash}, A. and {Bamba}, A. and {Baquero Larriva}, A. and {Baroncelli}, L. and {Barres de Almeida}, U. and {Barrio}, J.~A. and {Batkovic}, I. and {Baxter}, J. and {Becerra Gonz{\'a}lez}, J. and {Bernardini}, E. and {Bernardos}, M.~I. and {Bernete Medrano}, J. and {Berti}, A. and {Bhattacharjee}, P. and {Biederbeck}, N. and {Bigongiari}, C. and {Bissaldi}, E. and {Blanch}, O. and {Bordas}, P. and {Buisson}, C. and {Bulgarelli}, A. and {Burelli}, I. and {Buscemi}, M. and {Cardillo}, M. and {Caroff}, S. and {Carosi}, A. and {Cassol}, F. and {Cauz}, D. and {Ceribella}, G. and {Chai}, Y. and {Cheng}, K. and {Chiavassa}, A. and {Chikawa}, M. and {Chytka}, L. and {Cifuentes}, A. and {Contreras}, J.~L. and {Cortina}, J. and {Costantini}, H. and {D'Amico}, G. and {Dalchenko}, M. and {De Angelis}, A. and {de Bony de Lavergne}, M. and {De Lotto}, B. and {de Menezes}, R. and {Deleglise}, G. and {Delgado}, C. and {Delgado Mengual}, J. and {della Volpe}, D. and {Dellaiera}, M. and {Di Piano}, A. and {Di Pierro}, F. and {Di Tria}, R. and {Di Venere}, L. and {D{\'\i}az}, C. and {Dominik}, R.~M. and {Dominis Prester}, D. and {Donini}, A. and {Dorner}, D. and {Doro}, M. and {Els{\"a}sser}, D. and {Emery}, G. and {Escudero}, J. and {Fallah Ramazani}, V. and {Ferrara}, G. and {Fiasson}, A. and {Freixas Coromina}, L. and {Fr{\"o}se}, S. and {Fukami}, S. and {Fukazawa}, Y. and {Garcia}, E. and {Garcia L{\'o}pez}, R. and {Gasparrini}, D. and {Geyer}, D. and {Giesbrecht Paiva}, J. and {Giglietto}, N. and {Giordano}, F. and {Giro}, E. and {Gliwny}, P. and {Godinovic}, N. and {Grau}, R. and {Green}, D. and {Green}, J. and {Gunji}, S. and {Hackfeld}, J. and {Hadasch}, D. and {Hahn}, A. and {Hashiyama}, K. and {Hassan}, T. and {Hayashi}, K. and {Heckmann}, L. and {Heller}, M. and {Herrera Llorente}, J. and {Hirotani}, K. and {Hoffmann}, D. and {Horns}, D. and {Houles}, J. and {Hrabovsky}, M. and {Hrupec}, D. and {Hui}, D. and {H{\"u}tten}, M. and {Imazawa}, R. and {Inada}, T. and {Inome}, Y. and {Ioka}, K. and {Iori}, M. and {Ishio}, K. and {Iwamura}, Y. and {Jacquemont}, M. and {Jimenez Martinez}, I. and {Jurysek}, J. and {Kagaya}, M. and {Karas}, V. and {Katagiri}, H. and {Kataoka}, J. and {Kerszberg}, D. and {Kobayashi}, Y. and {Kong}, A. and {Kubo}, H. and {Kushida}, J. and {Lainez}, M. and {Lamanna}, G. and {Lamastra}, A. and {Le Flour}, T. and {Linhoff}, M. and {Longo}, F. and {L{\'o}pez-Coto}, R. and {L{\'o}pez-Moya}, M. and {L{\'o}pez-Oramas}, A. and {Loporchio}, S. and {Lorini}, A. and {Luque-Escamilla}, P.~L. and {Majumdar}, P. and {Makariev}, M. and {Mandat}, D. and {Manganaro}, M. and {Manic{\`o}}, G. and {Mannheim}, K. and {Mariotti}, M. and {Marquez}, P. and {Marsella}, G. and {Mart{\'\i}}, J. and {Martinez}, O. and {Mart{\'\i}nez}, G. and {Mart{\'\i}nez}, M. and {Marusevec}, P. and {Mas-Aguilar}, A. and {Maurin}, G. and {Mazin}, D. and {Mestre Guillen}, E. and {Micanovic}, S. and {Miceli}, D. and {Miener}, T. and {Miranda}, J.~M. and {Mirzoyan}, R. and {Mizuno}, T. and {Molero Gonzalez}, M. and {Molina}, E. and {Montaruli}, T. and {Monteiro}, I. and {Moralejo}, A. and {Morcuende}, D. and {Morselli}, A. and {Mrakovcic}, K. and {Murase}, K. and {Nagai}, A. and {Nakamori}, T. and {Nickel}, L. and {Nievas}, M. and {Nishijima}, K. and {Noda}, K. and {Nosek}, D. and {Nozaki}, S. and {Ohishi}, M. and {Ohtani}, Y. and {Okazaki}, N. and {Okumura}, A. and {Orito}, R. and {Otero-Santos}, J. and {Palatiello}, M. and {Paneque}, D. and {Pantaleo}, F.~R. and {Paoletti}, R. and {Paredes}, J.~M. and {Pavleti{\'c}}, L. and {Pech}, M. and {Pecimotika}, M. and {Pietropaolo}, E. and {Pirola}, G. and {Podobnik}, F. and {Poireau}, V. and {Polo}, M. and {Pons}, E. and {Prandini}, E. and {Prast}, J. and {Priyadarshi}, C. and {Prouza}, M. and {Rando}, R. and {Rhode}, W. and {Rib{\'o}}, M. and {Rizi}, V. and {Rodriguez Fernandez}, G. and {Saito}, T. and {Sakurai}, S. and {Sanchez}, D.~A. and {{\v{S}}ari{\'c}}, T. and {Saturni}, F.~G. and {Scherpenberg}, J. and {Schleicher}, B. and {Schmuckermaier}, F. and {Schubert}, J.~L. and {Schussler}, F. and {Schweizer}, T. and {Seglar Arroyo}, M. and {Sitarek}, J. and {Sliusar}, V. and {Spolon}, A. and {Stri{\v{s}}kovi{\'c}}, J. and {Strzys}, M. and {Suda}, Y. and {Sunada}, Y. and {Tajima}, H. and {Takahashi}, M. and {Takahashi}, H. and {Takata}, J. and {Takeishi}, R. and {Tam}, P.~H.~T. and {Tanaka}, S.~J. and {Tateishi}, D. and {Temnikov}, P. and {Terada}, Y. and {Terauchi}, K. and {Terzic}, T. and {Teshima}, M. and {Tluczykont}, M. and {Tokanai}, F. and {Torres}, D.~F. and {Travnicek}, P. and {Truzzi}, S. and {Tutone}, A. and {Uhlrich}, G. and {Vacula}, M. and {V{\'a}zquez Acosta}, M. and {Verguilov}, V. and {Viale}, I. and {Vigliano}, A. and {Vigorito}, C.~F. and {Vitale}, V. and {Voutsinas}, G. and {Vovk}, I. and {Vuillaume}, T. and {Walter}, R. and {Will}, M. and {Yamamoto}, T. and {Yamazaki}, R. and {Yoshida}, T. and {Yoshikoshi}, T. and {Zywucka (CTA-LST Project)}, N. and {Balbo}, M. and {Eckert}, D. and {Tramacere}, A.},
        title = "{Multiwavelength study of the galactic PeVatron candidate LHAASO J2108+5157}",
      journal = {\aap},
         year = 2023,
        month = may,
       volume = {673},
          eid = {A75},
        pages = {A75},
          doi = {10.1051/0004-6361/202245086},
archivePrefix = {arXiv},
       eprint = {2210.00775},
 primaryClass = {astro-ph.HE},
       adsurl = {https://ui.adsabs.harvard.edu/abs/2023A&A...673A..75A}
}

@ARTICLE{Dame_2001,
       author = {{Dame}, T.~M. and {Hartmann}, Dap and {Thaddeus}, P.},
        title = "{The Milky Way in Molecular Clouds: A New Complete CO Survey}",
      journal = {\apj},
         year = 2001,
        month = feb,
       volume = {547},
       number = {2},
        pages = {792-813},
          doi = {10.1086/318388},
archivePrefix = {arXiv},
       eprint = {astro-ph/0009217},
 primaryClass = {astro-ph},
       adsurl = {https://ui.adsabs.harvard.edu/abs/2001ApJ...547..792D}
}

@ARTICLE{Miville_2017,
       author = {{Miville-Desch{\^e}nes}, Marc-Antoine and {Murray}, Norman and {Lee}, Eve J.},
        title = "{Physical Properties of Molecular Clouds for the Entire Milky Way Disk}",
      journal = {\apj},
         year = 2017,
        month = jan,
       volume = {834},
       number = {1},
          eid = {57},
        pages = {57},
          doi = {10.3847/1538-4357/834/1/57},
archivePrefix = {arXiv},
       eprint = {1610.05918},
 primaryClass = {astro-ph.GA},
       adsurl = {https://ui.adsabs.harvard.edu/abs/2017ApJ...834...57M}
}

@ARTICLE{Fuente_2023a,
       author = {{de la Fuente}, Eduardo and {Toledano-Juarez}, Ivan and {Kawata}, Kazumasa and {Trinidad}, Miguel A. and {Tafoya}, Daniel and {Sano}, Hidetoshi and {Tokuda}, Kazuki and {Nishimura}, Atsushi and {Onishi}, Toshikazu and {Sako}, Takashi and {Hona}, Binita and {Ohnishi}, Munehiro and {Takita}, Masato},
        title = "{Detection of a new molecular cloud in the LHAASO J2108+5157 region supporting a hadronic PeVatron scenario}",
      journal = {\pasj},
         year = 2023,
        month = jun,
       volume = {75},
       number = {3},
        pages = {546-566},
          doi = {10.1093/pasj/psad018},
archivePrefix = {arXiv},
       eprint = {2303.05712},
 primaryClass = {astro-ph.HE},
       adsurl = {https://ui.adsabs.harvard.edu/abs/2023PASJ...75..546D}
}

@ARTICLE{Fuente_2023b,
       author = {{de la Fuente}, E. and {Toledano-Ju{\'a}rez}, I. and {Kawata}, K. and {Trinidad}, M.~A. and {Yamagishi}, M. and {Takekawa}, S. and {Tafoya}, D. and {Ohnishi}, M. and {Nishimura}, A. and {Kato}, S. and {Sako}, T. and {Takita}, M. and {Sano}, H. and {Yadav}, R.~K.},
        title = "{Evidence for a gamma-ray molecular target in the enigmatic PeVatron candidate LHAASO J2108+5157}",
      journal = {\aap},
         year = 2023,
        month = jul,
       volume = {675},
          eid = {L5},
        pages = {L5},
          doi = {10.1051/0004-6361/202346681},
archivePrefix = {arXiv},
       eprint = {2306.11921},
 primaryClass = {astro-ph.HE},
       adsurl = {https://ui.adsabs.harvard.edu/abs/2023A&A...675L...5D}
}

@ARTICLE{Abeysekara_2020,
       author = {{Abeysekara}, A.~U. and {Albert}, A. and {Alfaro}, R. and {Angeles Camacho}, J.~R. and {Arteaga-Vel{\'a}zquez}, J.~C. and {Arunbabu}, K.~P. and {Avila Rojas}, D. and {Ayala Solares}, H.~A. and {Baghmanyan}, V. and {Belmont-Moreno}, E. and {BenZvi}, S.~Y. and {Brisbois}, C. and {Caballero-Mora}, K.~S. and {Capistr{\'a}n}, T. and {Carrami{\~n}ana}, A. and {Casanova}, S. and {Cotti}, U. and {Cotzomi}, J. and {Couti{\~n}o de Le{\'o}n}, S. and {De la Fuente}, E. and {de Le{\'o}n}, C. and {Dichiara}, S. and {Dingus}, B.~L. and {DuVernois}, M.~A. and {D{\'\i}az-V{\'e}lez}, J.~C. and {Ellsworth}, R.~W. and {Engel}, K. and {Espinoza}, C. and {Fleischhack}, H. and {Fraija}, N. and {Galv{\'a}n-G{\'a}mez}, A. and {Garcia}, D. and {Garc{\'\i}a-Gonz{\'a}lez}, J.~A. and {Garfias}, F. and {Gonz{\'a}lez}, M.~M. and {Goodman}, J.~A. and {Harding}, J.~P. and {Hernandez}, S. and {Hinton}, J. and {Hona}, B. and {Huang}, D. and {Hueyotl-Zahuantitla}, F. and {H{\"u}ntemeyer}, P. and {Iriarte}, A. and {Jardin-Blicq}, A. and {Joshi}, V. and {Kaufmann}, S. and {Kieda}, D. and {Lara}, A. and {Lee}, W.~H. and {Le{\'o}n Vargas}, H. and {Linnemann}, J.~T. and {Longinotti}, A.~L. and {Luis-Raya}, G. and {Lundeen}, J. and {L{\'o}pez-Coto}, R. and {Malone}, K. and {Marinelli}, S.~S. and {Martinez}, O. and {Martinez-Castellanos}, I. and {Mart{\'\i}nez-Castro}, J. and {Mart{\'\i}nez-Huerta}, H. and {Matthews}, J.~A. and {Miranda-Romagnoli}, P. and {Morales-Soto}, J.~A. and {Moreno}, E. and {Mostaf{\'a}}, M. and {Nayerhoda}, A. and {Nellen}, L. and {Newbold}, M. and {Nisa}, M.~U. and {Noriega-Papaqui}, R. and {Peisker}, A. and {P{\'e}rez-P{\'e}rez}, E.~G. and {Pretz}, J. and {Ren}, Z. and {Rho}, C.~D. and {Rivi{\`e}re}, C. and {Rosa-Gonz{\'a}lez}, D. and {Rosenberg}, M. and {Ruiz-Velasco}, E. and {Salesa Greus}, F. and {Sandoval}, A. and {Schneider}, M. and {Schoorlemmer}, H. and {Sinnis}, G. and {Smith}, A.~J. and {Springer}, R.~W. and {Surajbali}, P. and {Tabachnick}, E. and {Tanner}, M. and {Tibolla}, O. and {Tollefson}, K. and {Torres}, I. and {Torres-Escobedo}, R. and {Villase{\~n}or}, L. and {Weisgarber}, T. and {Wood}, J. and {Yapici}, T. and {Zhang}, H. and {Zhou}, H. and {HAWC Collaboration}},
        title = "{Multiple Galactic Sources with Emission Above 56 TeV Detected by HAWC}",
      journal = {\prl},
         year = 2020,
        month = jan,
       volume = {124},
       number = {2},
          eid = {021102},
        pages = {021102},
          doi = {10.1103/PhysRevLett.124.021102},
archivePrefix = {arXiv},
       eprint = {1909.08609},
 primaryClass = {astro-ph.HE},
       adsurl = {https://ui.adsabs.harvard.edu/abs/2020PhRvL.124b1102A}
}

@ARTICLE{Albert_2020,
       author = {{Albert}, A. and {Alfaro}, R. and {Alvarez}, C. and {Camacho}, J.~R. Angeles and {Arteaga-Vel{\'a}zquez}, J.~C. and {Arunbabu}, K.~P. and {Avila Rojas}, D. and {Ayala Solares}, H.~A. and {Baghmanyan}, V. and {Belmont-Moreno}, E. and {BenZvi}, S.~Y. and {Brisbois}, C. and {Caballero-Mora}, K.~S. and {Capistr{\'a}n}, T. and {Carrami{\~n}ana}, A. and {Casanova}, S. and {Cotti}, U. and {Couti{\~n}o de Le{\'o}n}, S. and {De la Fuente}, E. and {Diaz Hernandez}, R. and {Diaz-Cruz}, L. and {Dingus}, B.~L. and {DuVernois}, M.~A. and {Durocher}, M. and {D{\'\i}az-V{\'e}lez}, J.~C. and {Ellsworth}, R.~W. and {Engel}, K. and {Espinoza}, C. and {Fan}, K.~L. and {Fang}, K. and {Alonso}, M. Fern{\'a}ndez and {Fleischhack}, H. and {Fraija}, N. and {Galv{\'a}n-G{\'a}mez}, A. and {Garcia}, D. and {Garc{\'\i}a-Gonz{\'a}lez}, J.~A. and {Garfias}, F. and {Giacinti}, G. and {Gonz{\'a}lez}, M.~M. and {Goodman}, J.~A. and {Harding}, J.~P. and {Hernandez}, S. and {Hinton}, J. and {Hona}, B. and {Huang}, D. and {Hueyotl-Zahuantitla}, F. and {H{\"u}ntemeyer}, P. and {Iriarte}, A. and {Jardin-Blicq}, A. and {Joshi}, V. and {Kieda}, D. and {Lara}, A. and {Lee}, W.~H. and {Le{\'o}n Vargas}, H. and {Linnemann}, J.~T. and {Longinotti}, A.~L. and {Luis-Raya}, G. and {Lundeen}, J. and {L{\'o}pez-Coto}, R. and {Malone}, K. and {Marandon}, V. and {Martinez}, O. and {Martinez-Castellanos}, I. and {Mart{\'\i}nez-Castro}, J. and {Matthews}, J.~A. and {Miranda-Romagnoli}, P. and {Morales-Soto}, J.~A. and {Moreno}, E. and {Mostaf{\'a}}, M. and {Nayerhoda}, A. and {Nellen}, L. and {Newbold}, M. and {Nisa}, M.~U. and {Noriega-Papaqui}, R. and {Olivera-Nieto}, L. and {Omodei}, N. and {Peisker}, A. and {P{\'e}rez Araujo}, Y. and {P{\'e}rez-P{\'e}rez}, E.~G. and {Ren}, Z. and {Rho}, C.~D. and {Rivi{\`e}re}, C. and {Rosa-Gonz{\'a}lez}, D. and {Ruiz-Velasco}, E. and {Salazar}, H. and {Salesa Greus}, F. and {Sandoval}, A. and {Schneider}, M. and {Schoorlemmer}, H. and {Serna}, F. and {Sinnis}, G. and {Smith}, A.~J. and {Springer}, R.~W. and {Surajbali}, P. and {Tollefson}, K. and {Torres}, I. and {Torres-Escobedo}, R. and {Ukwatta}, T.~N. and {Ure{\~n}a-Mena}, F. and {Weisgarber}, T. and {Werner}, F. and {Willox}, E. and {Zepeda}, A. and {Zhou}, H. and {de Le{\'o}n}, C. and {{\'A}lvarez}, J.~D. and {HAWC Collaboration}},
        title = "{3HWC: The Third HAWC Catalog of Very-high-energy Gamma-Ray Sources}",
      journal = {\apj},
         year = 2020,
        month = dec,
       volume = {905},
       number = {1},
          eid = {76},
        pages = {76},
          doi = {10.3847/1538-4357/abc2d8},
archivePrefix = {arXiv},
       eprint = {2007.08582},
 primaryClass = {astro-ph.HE},
       adsurl = {https://ui.adsabs.harvard.edu/abs/2020ApJ...905...76A}
}

@ARTICLE{Amenomori_2019,
       author = {{Amenomori}, M. and {Bao}, Y.~W. and {Bi}, X.~J. and {Chen}, D. and {Chen}, T.~L. and {Chen}, W.~Y. and {Chen}, Xu and {Chen}, Y. and {Cirennima} and {Cui}, S.~W. and {Danzengluobu} and {Ding}, L.~K. and {Fang}, J.~H. and {Fang}, K. and {Feng}, C.~F. and {Feng}, Zhaoyang and {Feng}, Z.~Y. and {Gao}, Qi and {Gou}, Q.~B. and {Guo}, Y.~Q. and {He}, H.~H. and {He}, Z.~T. and {Hibino}, K. and {Hotta}, N. and {Hu}, Haibing and {Hu}, H.~B. and {Huang}, J. and {Jia}, H.~Y. and {Jiang}, L. and {Jin}, H.~B. and {Kajino}, F. and {Kasahara}, K. and {Katayose}, Y. and {Kato}, C. and {Kato}, S. and {Kawata}, K. and {Kozai}, M. and {Labaciren} and {Le}, G.~M. and {Li}, A.~F. and {Li}, H.~J. and {Li}, W.~J. and {Lin}, Y.~H. and {Liu}, B. and {Liu}, C. and {Liu}, J.~S. and {Liu}, M.~Y. and {Lou}, Y. -Q. and {Lu}, H. and {Meng}, X.~R. and {Mitsui}, H. and {Munakata}, K. and {Nakamura}, Y. and {Nanjo}, H. and {Nishizawa}, M. and {Ohnishi}, M. and {Ohta}, I. and {Ozawa}, S. and {Qian}, X.~L. and {Qu}, X.~B. and {Saito}, T. and {Sakata}, M. and {Sako}, T.~K. and {Sengoku}, Y. and {Shao}, J. and {Shibata}, M. and {Shiomi}, A. and {Sugimoto}, H. and {Takita}, M. and {Tan}, Y.~H. and {Tateyama}, N. and {Torii}, S. and {Tsuchiya}, H. and {Udo}, S. and {Wang}, H. and {Wu}, H.~R. and {Xue}, L. and {Yagisawa}, K. and {Yamamoto}, Y. and {Yang}, Z. and {Yuan}, A.~F. and {Zhai}, L.~M. and {Zhang}, H.~M. and {Zhang}, J.~L. and {Zhang}, X. and {Zhang}, X.~Y. and {Zhang}, Y. and {Zhang}, Yi and {Zhang}, Ying and {Zhaxisangzhu} and {Zhou}, X.~X. and {Tibet AS {\ensuremath{\gamma}} Collaboration}},
        title = "{First Detection of Photons with Energy beyond 100 TeV from an Astrophysical Source}",
      journal = {\prl},
         year = 2019,
        month = aug,
       volume = {123},
       number = {5},
          eid = {051101},
        pages = {051101},
          doi = {10.1103/PhysRevLett.123.051101},
archivePrefix = {arXiv},
       eprint = {1906.05521},
 primaryClass = {astro-ph.HE},
       adsurl = {https://ui.adsabs.harvard.edu/abs/2019PhRvL.123e1101A}
}

@ARTICLE{Eikenberry_2001,
       author = {{Eikenberry}, S.~S. and {Fischer}, W.~J. and {Egami}, E. and {Djorgovski}, S.~G.},
        title = "{Deep Infrared Imaging of the Microquasars 1E 1740-2942 and GRS 1758-258}",
      journal = {\apj},
         year = 2001,
        month = jul,
       volume = {556},
       number = {1},
        pages = {1-5},
          doi = {10.1086/321560},
archivePrefix = {arXiv},
       eprint = {astro-ph/0103422},
 primaryClass = {astro-ph},
       adsurl = {https://ui.adsabs.harvard.edu/abs/2001ApJ...556....1E}
}

@ARTICLE{Urry_1995,
       author = {{Urry}, C. Megan and {Padovani}, Paolo},
        title = "{Unified Schemes for Radio-Loud Active Galactic Nuclei}",
      journal = {\pasp},
         year = 1995,
        month = sep,
       volume = {107},
        pages = {803},
          doi = {10.1086/133630},
archivePrefix = {arXiv},
       eprint = {astro-ph/9506063},
 primaryClass = {astro-ph},
       adsurl = {https://ui.adsabs.harvard.edu/abs/1995PASP..107..803U}
}
\bibliographystyle{aasjournalv7}



\end{document}